\documentstyle[a4,cite,12pt,aps,epsfig]{revtex}

\makeatletter
\def\dgr{\dagger}
\def\nnb{\nonumber}
\newcommand{\bea}{\begin{eqnarray}}
\newcommand{\eea}{\end{eqnarray}}

\makeatother

\begin{document}

\draft

\title{Is the truncated SU(N) non-Abelian gauge theory in extra dimensions renormalizable?}

\author{
             {\bf Qi-Shu Yan$^b$}\footnote{
        E-mail : yanqs@mail.ihep.ac.cn} and {\bf Dong-Sheng Du$^a,^b$}
\footnote{
        E-mail : dsdu@mail.ihep.ac.cn}  \\
       {\small\em a.  CCAST (World Laboratory), P.O.Box 8730, Beijing 100080, P.R.China\\}
       {\small\em b. IHEP of CAS, P.O.Box 918(4), Beijing 100039, P.R.China}
}
\bigskip

\address{\hfill{}}

\maketitle

\begin{abstract}
In this letter we show that in the extra dimension model,
contrary to the widely accepted conception,
the simply truncated $\phi^4$ and non-Abelian SU(N) Kaluza-Klein
theories are not renormalizable, i.e. the tree level relations of the
effective theories can not sustain the quantum corrections.
The breaking down of the tree level relations of
the effective theories can be traced back to several factors:
the breaking of the higher dimension Lorentz symmetry and
higher dimension gauge symmetry, interactions assumed in the underlying
Lagrangians, and the dimension reduction and rescaling procedure.
\end{abstract}

Renormalization holds a quite special
role in the development of the quantum field
theory \cite{thooft1}.
As we know, quantum corrections of the 4D quantum field
theory are generally infinite, and only in a
renormalizable theory is it possible
through the standard renormalization procedure to remove
the ultraviolet divergences in the theory by
introducing only few finite counter
terms and to make loop contributions (quantum corrections)
finite and meaningful.

By considering the degrees of superficial divergence
of the irreducible vertices of a specified quantum field
theory defined in D dimension, the criterion of renormalizability
can be simply formulated \cite{dyson, weinberg} as
\bea
\Omega = D - \sum_{i=1}^{n} d_i- \frac{D-1}{2} E_f - \frac{D-2}{2}E_b,
\label{critiria}
\eea
where $\Omega$ is the superficial divergence
of any a Feynman integral determined by the theory,
$d_i$ is the mass dimension of couplings of the theory,
and $E_f$ ($E_b$) is the number of external
fermions (bosons). This equation tells us
that a theory with couplings of positive or vanishing mass
dimension is (super-)renormalizable, while a theory with
couplings of negative mass dimension is non-renormalizable.
And the non-renormalizability of a quantum field theory in
extra dimension becomes a straightforward inference due to
the fact that any a couplings in the theory (except the $\phi^3$
in 5D and 6D \cite{collins}) will have a negative mass dimension.

In the non-Abelian gauge theory, we meet another kind of
problem of renormalizability. The theory is unquestionably
renormalizable if only judged from the power law given in
Eq. (\ref{critiria}). But it is not sufficient. In the pure
Yang-Mills theory for instance, there is only one
coupling constant in the theory which determines
both the trilinear and quartic couplings of vector bosons
and the ghost-ghost-vector coupling, as required by the
quantum gauge covariance. A subtle problem
arises: whether the tree-level gauge structure preserves
after taking into account the quantum corrections. In
another word, whether the counter term determined by, say,
the three point Green function, is enough to eliminate the
ultraviolet divergences of the four point Green function of
vector boson and the ghost-ghost-vector interaction. As we
know, the BRST symmetry \cite{brst} and the Slavnov-Taylor
identities \cite{st} guarantee the tree level
gauge structure of the theory order by order, and
the pure Yang-Mills gauge theory is renormalizable \cite{thooft2}.
When more particles are added to a non-Abelian
gauge theory, if there is no anomaly, we know the theory is
still renormalizable, even in the case
when the gauge symmetry is spontaneously broken.

The extra dimension theory is a fast developing topic
in recent years, and two kinds of extra dimensions can be roughly
divided: large extra dimensions where gravity is considered,
and small extra dimensions where the standard model is extended to
the high dimensions. We concern the later case here, and there are
many papers on both models and phenomenologies of it \cite{ddg, phkk}.
But, there is an irksome problem about it is that theoretical
predictions are explicitly cutoff-dependent even in
tree level calculations due to the sum of infinite Kaluza-Klein
(KK) excitations, such a fact can be traced
back to the intrinsic non-renormalizability of the
higher dimension quantum theory. Furthermore, the trouble
becomes even more serious for the loop processes.

There are papers to regularize the divergent contribution of
KK excitations \cite{rkk}, and it seems only the string
regularization can provide a solid solution to the problem
\cite{strreg}. Recently, the renormalizable
effective theory of the extra dimension is constructed
in reference \cite{decon}, where the mass generation
mechanism of the compactification of extra dimension
is non-linearly realized in a technicolor way or in the
latticed extra dimension.
The (de)constructing way only provides an effective description
of the extra dimension theory, but doesn't prove that an extra
dimension theory (or a simply truncated theory) is renormalizable.

To evaluate the contribution of KK excitations,
a widely accepted and practical conception
indicated in the literatures is to truncate
the infinite KK towers to finite. With the belief
that the truncated KK theories are always renormalizable, the
tree-level relations among couplings are always used to make
theoretical predictions, both in tree level and one-loop level.
However in this letter we will show that the tree level
relations of the effective theory might be broken by the
quantum corrections. Considering the characteristic power
running of extra dimension models, a large deviation from
the tree level relations might be caused, therefore from either
the theoretical respect or the numerical and practical respect,
this conception is quite questionable.
Below we will detail this problem in two cases: the $\phi^4$ theory
and non-Abelian SU(N) gauge theory defined in 5D. In order to contrast
and compare, we will also examine the QED and $\phi^3$ theory in 5D.

We examine the $\phi^4$ theory first.
The Lagrangian of the $\phi^4$ theory in 5D is defined as
\bea
L=(\partial_{M} \phi_{5D})^{\dgr} (\partial^{M} \phi_{5D})
-m^2 (\phi_{5D})^{\dgr} \phi_{5D} - {\lambda_{5D} \over 4} \left((\phi_{5D})^{\dgr} \phi_{5D}\right)^2,
\label{lag}
\eea
where $M=0,1,2,3,5$. The complex singlet field $\phi_{5D}$ and quartic
coupling $\lambda_{5D}$ have
the mass dimensions $3/2$ and $-1$, respectively.
This Lagrangian owns a 5D Lorentz
space-time symmetry and global U(1) inner symmetry with
the universal phase defined in 5D.

And according to the power law given in Eq. (\ref{critiria}),
this theory is non-renormalizable.
However, it is helpful to understand the Lagrangian given in
Eq. (\ref{lag}) in Wilson's renormalization method \cite{weinberg},
which is valid for quantum field theories defined in any dimension of
spacetime. In this method, the principle of renormalizability is not
necessary. The price paid for the sacrifice of this restrictive principle
is that one has to include all interactions in the effective Lagrangian
permitted by the 5D spacetime Lorentz and 5D gauge symmetry, and the number
of these operators is infinite. In the $\phi^4$ case we consider here,
besides the minimal interaction term $(\phi^{\dgr} \phi)^2$, interactions
like $(\phi^{\dgr} \phi)^3$, $\phi^{\dgr} \Box^2 \phi$, etc. should also
be added to the Lagrangian given in Eq. (\ref{lag}). According to the
effective theory \cite{eqft}, at low energy region the interactions
with lower dimensions domain. So the Lagrangian given in
Eq. (\ref{lag}) can only be understood as being valid below a given
ultraviolet cutoff $\Lambda_{UV}^{5D}$, where operators with higher
dimensions have been greatly suppressed.
Therefore the Lagrangian given in Eq. (\ref{lag})
should be only valid for $|P_{5D}| < \Lambda_{UV}^{5D}$, otherwise the
unitarity of the S-matrix will be violated if $|P_{5D}|$ is much greater
than $\Lambda_{UV}^{5D}$
(Here $|P_{5D}| =\sqrt{p_M^2}, M=0,1,2,3,5$, the metric of spacetime is
taken as that of a Ecludian one.).

The Lagrangian given in Eq. (\ref{lag})
has also an infrared cutoff $\Lambda_{IR}^{5D}$ in the compactified extra dimension
theories when $\Lambda_{IR}^{5D}$ approaches the compactification scale $1/R_C$
($R_C$ is the compactification size).
The reason for this infrared cutoff is that near the energy region $1/R_C$,
it would be not appropriate any longer to regard the fifth dimension as
infinite large and use the 5D Lorentz symmetry and 5D gauge symmetry to restrict
operators which might appear in its effective Lagrangian.

For the small extra dimension scenarios,
the extra dimensions are always
assumed to by compactified and small (say TeV size).
In order to match with the low energy regions
where the observed world is 4D,
the standard dimension reduction method
and the matching procedure are used to derive the effective 4D quantum field theory.
For example, by assuming that the vacuum manifold has
a $M_4\times S^1/Z_2$ structure (the 5D Lorentz space-time
symmetry is broken by the vacuum while the U(1) symmetry
should also be modified), and by requiring
that the Lagrangian is invariant under the
orbifold transformation $x_5 \rightarrow -x_5$,
we can assign a boundary condition for the $\phi_{5D}$:
$\phi_{5D}(x,x_5)=-\phi_{5D}(x,-x_5)$.
Then $\phi_{5D}$ field can be Fourier-expanded as
\bea
\phi_{5D}(x, x_5)=\phi_{5D}^{n} \cos\frac{n x_5}{R_c}.
\label{decomp}
\eea
Substituting the Eq. (\ref{decomp}) into the Lagrangian given
in the Eq. (\ref{lag}) and integrating out the fifth component
of the space-time, we get the following
reduced effective 4D theory (RE4DT)
\bea
L^{eff}&=&L_{kin}
+ L_{int}\\
L_{kin}&=&\sum_{n=0}\phi^{n\dgr}\left(-\partial^{\mu} \partial_{\mu}-\frac{n^2}{R_c^2}-m^2\right) \phi^{n}\\
L_{int}&=&-{\lambda \over 4} \left\{ \left(\phi^{0\dgr} \phi^{0}\right)^2
+ \sum_{k,l,m=1}^{\infty} R_1(k,l,m) \left( \phi^{0\dgr} \phi^{k} \phi^{l\dgr} \phi^{m\dgr} + h.c.\right) \right.\nnb\\
&&\left.+ \sum_{n=1}^{\infty} \left[ 4 \phi^{0\dgr} \phi^{0} \phi^{n\dgr} \phi^{n} +
2 Re\left(\phi^{0\dgr} \phi^{n} \phi^{0\dgr} \phi^{n}\right) \right] \right.\nnb\\
&&\left.+ \sum_{k,l,m,n=1}^{\infty} R_2(k,l,m,n) \phi^{k\dgr} \phi^{l} \phi^{m\dgr} \phi^{n} \right \}
\label{eff}
\eea
where $R_{i}, i=1,\,\,2$ are normalization factors and can
be understood as the requirement of the momentum conservation
of the fifth dimension.
Here we omit the subscript 4D for all quantities.
To get the RE4DT, the following rescaling relations has been used
\bea
\phi^{0}_{4D} \to \sqrt{2 \pi R_c} \phi^{0}_{5D},
\phi^{n}_{4D} \to \sqrt{\pi R_c}\phi^{n}_{5D},
\lambda_{4D}  \to {\lambda \over {2 \pi R_c}}.
\eea
The theory owns a 4D space-time symmetry and the reduced global U(1)
symmetry. The RE4DT is invariant under the following transformation
\bea
\phi^{n} \rightarrow exp(i \alpha) \phi^{n}\,\,.
\eea

It is remarkable that there is an infinite KK towers in the theory,
and the zero modes have a different
normalization factor than the other KK excitations.
Another remarkable fact is that the infinite interactions
among KK modes are controlled by only one parameter $\lambda$.

Now the effects of high dimension are effectively
reflected by the infinite KK towers appeared in
the RE4DT given in the Eq. (\ref{eff}).
There are no coupling which has negative mass dimension in the
theory, and from the power law,
it seems that the theory should be renormalizable and
the dimension reduction procedure makes a higher dimension
theory to a renormalizable one.
But, due to the infinite KK excitations,
even if the contribution to a process of each KK excitation
is finite, the total result might still be infinite.
In this sense, the RE4DT is still non-renormalizable.

To effectively describe the 5D theory given in Eq. (\ref{lag}),
we must match its RE4DT with the underlying 5D theory at
a given scale $\Lambda'$, which should be in the range
$\Lambda_{IR}^{5D} < \Lambda < \Lambda_{UV}^{5D}$.
Therefore, the infinite KK excitations are truncated
by requiring $N'/R_C \approx \Lambda'$ ($N'/R_C$ is the heaviest
KK excitation included in the RE4DT $L_{\Lambda'}^{4D}$)
and only finite KK excitations
are kept in the RE4DT $L_{\Lambda'}^{4D}$.
Then finite results could be obtained even for loop
processes. It is in this sense the truncated KK theory
is renormalizable.

But is that all? Since the couplings among KK modes are
controlled by only one parameter $\lambda_{4D}$, then it is
naturally to ask: whether is it enough to introduce just
only one counter term to eliminate all ultraviolet divergences
in the effective theory? Or in other words, can the tree level
structure sustain the quantum corrections?
The problem is quite similar to the case for
the non-Abelian gauge theory in 4D.

In the underlying 5D theory, the answer to this problem
is affirmative. To demonstrate the reason, let's consider
to match the RE4DT with the underlying 5D theory at another
scale $\Lambda''$, and for the sake of convenience, we
assume that $\Lambda_{IR}^{5D} < \Lambda''< \Lambda' < \Lambda_{UV}^{5D}$.
So after invoking the matching
procedure at $\Lambda''$, we will get the $L_{\Lambda''}^{4D}$ with
$N''$ KK excitations ($N''$ is determined by $N''/R_C \approx \Lambda''$).
There are two differences between the $L_{\Lambda'}^{4D}$
and $L_{\Lambda''}^{4D}$:
1) the numbers of KK excitations are different, the $L_{\Lambda''}^{4D}$ can
be obtained by successively integrating out $N'-N''$ KK excitations;
2) the values of couplings $\lambda_{5D}(\Lambda'')$ and $\lambda_{5D}(\Lambda')$
are different, but are related with each other by the
renormalization group equation (RGE) of $\lambda_{5D}$.
However, there is a common between these two RE4DTs: the tree-level relations
among KK excitations seem to be hold.
Since the RGE is valid in loop level, then it might
tantalize one to expect that these tree-level relations would
also hold in the RE4DTs in loop-level. However, we will show
that it's not the case!

To simplify consideration, we truncate the infinite KK
excitations and keep only the $0-$ and $1-$
modes in the RE4DT. In order to find the consistent
solution to the requirement of renormalizability,
we rewrite the interaction part
of the Lagrangian in a more general form
\bea
- L_{int}&=&{\lambda_{00} \over 4} (\phi^{0\dgr} \phi^{0})^2 + {\lambda_{11} \over 4} (\phi^{1\dgr} \phi^{1})^2
+ {\lambda_{01} \over4 } \left [ 4 (\phi^{0\dgr} \phi^{0})(\phi^{1\dgr} \phi^{1})
+ 2 Re(\phi^{0\dgr} \phi^{1} \phi^{0\dgr} \phi^{1})\right ].
\eea
The RE4DT is only a special case of the interaction and gives
\bea
R \lambda=R \lambda_{00}=R \lambda_{01}=\lambda_{11},
\label{init}
\eea
where $R=3/2$. Now we determine the counter terms of the theory.
The counter terms, $\delta\lambda_{00}$,
$\delta\lambda_{01}$, and $\delta\lambda_{11}$ of
$\lambda_{00}$, $\lambda_{01}$, and $\lambda_{11}$,
can be directly constructed from the one loop diagrams.
In the dimension regularization and
${\bar{MS}}$ renormalization scheme,
the $\delta\lambda_{00}$, $\delta\lambda_{01}$,
and $\delta\lambda_{11}$ are simply determined as
\bea
\delta\lambda_{00}&=&{3\over 2} \kappa \Delta_{\epsilon} (\lambda_{00}^2 + \lambda_{01}^2)\\
\delta\lambda_{01}&=&{1\over 2} \kappa \Delta_{\epsilon} \left( \lambda_{01} \lambda_{00} + \lambda_{01} \lambda_{11} + 4 \lambda_{01}^2\right )\\
\delta\lambda_{11}&=&{3\over 2} \kappa \Delta_{\epsilon} (\lambda_{11}^2 + \lambda_{01}^2)
\label{dell}
\eea
where $\kappa=1/(16 \pi^2)$,
$\Delta_{\epsilon}=2/\epsilon-\gamma_{E}+log4\pi$, and
$\epsilon=4-D$. With these counter terms, the consistent
solution can be easily found. If the RE4DT is renormalizable,
we hope that the following relation should hold
\bea
\delta\lambda_{00}=\delta\lambda_{01}=\delta\lambda_{11},
\label{condition}
\eea
then the consistent solution for this equation requires
\bea
\lambda_{00}=\lambda_{01}=\lambda_{11}
\label{sol}
\eea
But the tree level relation given in the
Eq. (\ref{init}) obviously isn't
satisfying Eq. (\ref{sol}). Therefore it
is impossible to just introduce
one counter term $\delta\lambda$ to make the
quantum corrections of the theory finite, and
the tree level relation Eq. (\ref{init}) breaks
down. And it is in this sense that the RE4DT is still
non-renormalizable. For the truncated theory with more
than one KK excitations, we have the same conclusion.

It is remarkable that from the Eq. (\ref{dell}) we
know the tree level relation
$\lambda_{00}=\lambda_{01}$ will
also be broken down due to the
contribution from $\lambda_{11}$, so
it is questionable to use the relation
at low energy regions when evaluating
the contributions of KK excitations to
the effective potential of $\phi^0$.

Of course, if we forget the dimension reduction and adjust
the normalization factor $R$ to be just one,
then it is enough to just introduce
one counter term $\delta\lambda$ to make the
quantum corrections of the theory finite, at least up to one-loop.
Obviously, the procedure of normalizing
and rescaling in the standard dimension reduction,
which makes zero modes different from other
KK excitations and produces the normalization
factor $R_{i}$, is blamed for the non-renormalizability
of the theory. So we conclude here that the
non-renormalizability of the high dimension $\phi^4$
theory leaves its trace not only in appearing
the infinite KK excitations but in breaking down
the tree level relations among couplings with
quantum corrections. We also see here that the
reduced U(1) symmetry of the theory has no much
help on the problem in hand.

Equipped with this experience, it is naturally to ask
whether the tree level relations of the
truncated SU(N) gauge theory can sustain the quantum corrections.
Now we consider the case of non-Abelian SU(N) gauge theory.
The Lagrangian in 5D is given as
\bea
{\cal L} ~=~ -{1\over 4}\, F_{MN}F^{MN}
- {1\over {2 \xi}} F^2(A_M) + {\bar c} {{\delta F(A_M)}\over {\delta \alpha}} c,
\label{sunlag}
\eea
where $F_{MN}=\partial_{M} A_N - \partial_{N} A_M + f A_M A_N$,
and $f$ is the structure constant of the Lie algebra.
And $F(A_M)$ is the gauge fixing term and can be assumed \cite{ddg} to
have the form
\bea
F(A_M)=\partial_{M} A^{M}
\label{gfix}
\eea
The theory owns the Lorentz symmetry of 5D space-time and
BRST symmetry in 5D. But the theory is non-renormalizable
even only judge from the naive power law, since the gauge
coupling owns a negative mass dimension. So formally, 
even though the theory owns a gauge symmetry (BRST symmetry in 5D),
it is still non-renormalizable.

Similar to the argument in the $\phi^4$ theory in 5D, the
Lagrangian given in Eq. (\ref{sunlag}) can only be understood as
being valid below the ultraviolet cutoff $\Lambda^{5D}_{UV}$, otherwise
effects of other higher dimension operators will be important or
the unitarity condition of the S-matrix will be violated.

The vacuum manifold is assumed to have the structure
$M_4 \times S^1/Z_2$ and the Lorentz symmetry
of 5D is spontaneously broken.
Considering the fact that the 5D space-time symmetry is broken to
4D space-time symmetry, and the 5D gauge symmetry is broken to
4D gauge symmetry, below we will choose the gauge fixing term
\bea
F(A_M)=\partial_{\mu} A^{\mu} - \xi \partial_5 A^5.
\eea
The advantage to choose this gauge fixing term than the
one given in Eq. (\ref{gfix}) is that physical observables
are gauge parameter independent \footnote{The reference \cite{mpr} also
used this gauge fixing term.}.

By assigning
a boundary condition for the vector gauge field
\bea
A_{\mu}(x,x_5)= A_{\mu}(x,-x_5),
\eea
and decomposing quantum fields in 5D with
$A_{\mu}(x)=A_{\mu}^{n}(x) \cos\frac{nx_5}{R_c}$,
we get the RE4DT in the below form:
\bea
L^{eff}_{4D}=L^{00}+L^{ED},\,\,L^{ED}=L^{ED}_{kin} + L^{ED}_{int},
\eea
\bea
L^{ED}_{int}=L^{ED}_{K0}+L^{ED}_{KK},\,\,L^{ED}_{K0}=L^{ED}_{K0,tri} + L^{ED}_{K0,qua},
\eea
\bea
L^{ED}_{kin}&=&A_{\mu}^{0} \left(g^{\mu \nu} \partial^{\gamma} \partial_{\gamma} - \partial^{\mu} \partial^{\nu}(1-{1\over \xi})\right) A_{\nu}^{0}
+{\bar c^{0}} \left(- \partial^{\mu} \partial_{\mu}\right) c^{0}\nnb \\
&&+\sum_{n=1}^{\infty}{1\over 2} A_{\mu}^{n} \left(g^{\mu \nu} \partial^{\gamma} \partial_{\gamma} + g^{\mu \nu} \frac{n^2}{R_c^2} - \partial^{\mu} \partial^{\nu}(1-{1\over \xi})\right) A_{\nu}^{n}\nnb\\
&&+\sum_{n=1}^{\infty} {1 \over 2} A_5^{n}\left(-\partial^{\mu} \partial_{\mu} - \xi {n^2 \over R_c^2} \right)A_5^{n}
+\sum_{n=1}^{\infty} {\bar c^{n}} \left(- \partial^{\mu} \partial_{\mu} - \xi {n^2 \over R_c^2}\right) c^{n},
\eea
\bea
L^{ED}_{K0,tri}&=&-{1\over 2} g f^{abc}\sum_{n=1}^{\infty} \left  ( W^{0a\mu\nu} A_{\mu}^{nb} A_{\nu}^{n c} + 2 A_{\mu}^{0a} A_{\nu}^{n b} W^{n c\mu \nu}\right)\nnb\\
&&+g f^{abc} \sum_{n=1}^{\infty} A_{\mu}^{a0}A_5^{n b}\left(\partial^{\mu} A_5^{n c} + {n\over R_c} A^{n c\mu}\right)
+g f^{abc} \sum_{n=1}^{\infty} \partial^{\mu} {\bar c}^{n a} A_{\mu}^{0 b} c^{n c}.
\eea
Where $L^{00}$ represents terms of pure zero modes,
$L^{ED}_{K0, qua}$ represents the quartic coupling between the
zero and KK modes, and $L^{ED}_{KK}$ represents couplings
among KK excitations. Here we omit those interactions among KK excitations.
The Lagrangian owns a 4D Lorentz space-time symmetry
and the reduced BRST symmetry. There is a conservation law
of the fifth momentum, which can be viewed as the result from the
compactification of the fifth dimension space.
Again, it is remarkable that there is a infinite KK towers in the theory,
and the zero modes have a different
normalization factor than the other KK excitations.
And the infinite interactions
among KK modes are controlled by only one parameter $g$, the gauge coupling
constant.

The matching procedure will truncate the infinite KK excitations to
finite. And the tree level relations among couplings of
KK modes are expected to hold if one judges from the underlying
theory with the 5D Lorentz spacetime symmetry and 5D gauge symmetry.

In order to examine the renormalizability of the truncated
theory, as done in the $\phi^4$ case,
we truncate the infinite KK towers and keep only the $0-$ 
and $1-$ modes in the Lagrangian. And the Lagrangian has the following
form
\bea
L=L_{kin} + L_{int},\,\,L_{int}=L_{tri}+L_{qua},
\eea
\bea
L_{kin}&=&A_{\mu}^{0} \left(g^{\mu \nu} \partial^{\gamma} \partial_{\gamma}
 - \partial^{\mu} \partial^{\nu}(1-{1\over \xi})\right) A_{\nu}^{0}
+{\bar c^{0}} \left(- \partial^{\mu} \partial_{\mu}\right) c^{0}\nnb \\
&&+{1\over 2} A_{\mu}^{1} \left(g^{\mu \nu} \partial^{\gamma} \partial_{\gamma}
+ g^{\mu \nu} \frac{1}{R_c^2} - \partial^{\mu} \partial^{\nu}(1-{1\over \xi})\right) A_{\nu}^{1}\nnb\\
&&+{1 \over 2} A_5^{1}\left(-\partial^{\mu} \partial_{\mu} - \xi {1 \over R_c^2}\right) A_5^{1}
+{\bar c^{1}} \left(- \partial^{\mu} \partial_{\mu} - \xi {1 \over R_c^2}\right) c^{1},
\eea
\bea
L_{tri}&=&g f^{abc} \left \{ -{1\over2} (\partial_{\mu} A^{0a}_{\nu}
-\partial_{\nu} A^{0a}_{\mu}) A^{0 b\mu } A^{0 c\nu }
-{1\over2} (\partial_{\mu} A^{0a}_{\nu}-\partial_{\nu} A^{0a}_{\mu}) A^{1b\mu} A^{1c\nu} \right. \nnb\\
&&\left .-{1\over2} (\partial_{\mu} A^{1a}_{\nu}-\partial_{\nu} A^{1a}_{\mu})
(A^{0b\mu} A^{1c\nu}+A^{0b\mu} A^{1c\nu})
+\partial^{\mu} {\bar c^{0a}} A^{0b}_{\mu} c^{0c}+\partial^{\mu} {\bar c^{1a}} A^{0b}_{\mu} c^{1c} \right .\nnb\\
&&\left. +\partial^{\mu} {\bar c^{0a}} A^{1b}_{\mu} c^{1c}+\partial^{\mu} {\bar c^{1a}} A^{1b}_{\mu} c^{0c}
+{1\over R_c} A^{1 a\mu} A^{0b}_{\mu} A^{1c}_5
+{\xi \over R_c} {\bar c^{1a}} A^{1b}_{5} c^{0c}
+\partial^{\mu} A^{1a}_5 A^{0b}_{\mu} A^{1c}_5 \right \},
\eea
\bea
L_{qua}&=& g^2 f^{abe} f^{cde} \left \{
-{1\over 4} A^{0a}_{\mu} A^{0b}_{\nu} A^{0c\mu} A^{0d\nu}
-{1\over 2} A^{0a}_{\mu} A^{0b}_{\nu} A^{1c\mu} A^{1d\nu}\right. \nnb\\
&&\left. -{1\over 2} A^{0a}_{\mu} A^{1b}_{\nu} A^{0c\mu} A^{1d\nu}
-{1\over 2} A^{0a}_{\mu} A^{1b}_{\nu} A^{1c\mu} A^{0d\nu}
+{1\over 2} A^{0a}_{\mu} A^{1b}_5 A^{0c\mu} A^{d1}_5 \right.\nnb\\
&&\left.+{R_1\over 2} A^{1a}_{\mu} A^{1a}_5 A^{1c\mu} A^{d1}_5
- {R_2\over 4} A^{1a}_{\mu} A^{1b}_{\nu} A^{1c\mu} A^{1d\nu}
\right \},
\eea
where $R_1=1/2$ and $R_2=3/2$.

This simplified RE4DT has five particles, where
massless zero modes include $A^0_{\mu}$ and $c^0$
and the massive first KK excitation includes $A^1_{\mu}$,
$c^1$ and $A_5$. There are nine trilinear and
five quartic couplings, all are controlled just by
one coupling constant $g$. Generally, in the framework of
effective theory, we have only 4D spacetime Lorentz symmetry
and 4D SU(N) gauge symmetry of zero mode to restrict permitted
operators in the Lagrangian, and each of these couplings
might be treated as a free parameter, as we do in the $\phi^4$ case.
Besides, there might be some extra interactions like
$A^1_5 A^1_5 A^1_5 A^1_5$, which is still renormalizable in 4D
and is expected to play an important role in low
energy region. However, for the sake of simplicity, we use these
tree level relations to calculate and check whether these relations
are consistent with the requirement of renormalizability.

In order to simplify the discussion, we omit
the renormalization of mass and gauge terms,
and only consider the counter term of
the relevant vertices given below
\bea
{\delta L_{int}}&=& {\delta Z_{000}} A^{0a}_{\mu} A^{0b}_{\nu} A^{0c}_{\rho}
+{\delta Z_{011}} A^{0a}_{\mu} A^{1b}_{\nu} A^{1c}_{\rho}
+{\delta Z_{0000}} A^{0a}_{\mu} A^{0b}_{\nu} A^{0c}_{\rho} A^{0d}_{\sigma}\nnb\\
&&+{\delta Z_{0011}} A^{0a}_{\mu} A^{0b}_{\nu} A^{1c}_{\rho} A^{1d}_{\sigma}
+{\delta Z_{1111}} A^{1a}_{\mu} A^{1b}_{\nu} A^{1c}_{\rho} A^{1d}_{\sigma}.
\eea

If the theory were renormalizable (the tree level relations held),
these counter terms should have their structures as given below
\bea
{\delta Z_{000(011)}}&=&c_{000(011)} \,\, V_3,\\
{\delta Z_{(0000),(0011,1111)}}&=&c_{0000,(0011,1111)}\,\,V_4,
\eea
where $c_{i}$ should be number, and $V_3$ and $V_4$
have the below forms
\bea
V_3&=&g f^{abc} \left [ g^{\mu\nu}(p-q)^{\rho}
+ g^{\nu\rho} (q-k)^{\mu} + g^{\rho\nu} (k-p)^{\nu}\right ],\\
V_4&=&-i g^2 \left [f^{abe} f^{cde} (g^{\mu \rho} g^{\nu \sigma}
-g^{\mu \sigma} g^{\nu \rho})+f^{ace} f^{dbe} (g^{\mu \nu}
g^{\rho \sigma} -g^{\mu \sigma} g^{\nu \rho}) \right .\nnb \\
&&\left . +f^{ade} f^{bce} (g^{\mu \nu}
g^{\rho \sigma} -g^{\mu \rho} g^{\nu \sigma}) \right ]\nnb\\
&=&-i g^2 \left [g^{\mu \rho} g^{\nu \sigma} \Sigma^{ac}_{bd}
+ g^{\mu \sigma} g^{\nu \rho} \Sigma^{ad}_{bc}
+ g^{\mu \nu} g^{\rho \sigma} \Sigma^{ab}_{cd}\right ].
\eea
where $\Sigma^{ab}_{cd}=f^{ace} f^{bde} + f^{ade} f^{bce}$,
and $\Sigma^{ab}_{cd}$ is unchanged (symmetric) when
the indices $a(c)$ and $b(d)$, and (ab) and (cd)
interchange with each other.

And if the tree level relations among vertices were preserved
after considering the quantum corrections, relations given
below should also hold
\bea
Z_{000}^2 = Z_{A^0} Z_{0000},\,\,Z_{011} = {Z_{A^1} \over  Z_{A^0}}Z_{000},\\
Z_{0011}={Z_{A^1} \over Z_{A^0}} Z_{0000},\,\,Z_{1111}={Z_{A^1}^2 \over Z_{A^0}^2} R_2 Z_{0000},
\eea
where $Z_{A^0}$ and $Z_{A^1}$ are the renormalization constants
of wave-functions, $Z_{000}$, $Z_{011}$, $Z_{0000}$, $Z_{0011}$,
and $Z_{1111}$ are the renormalization constants of the corresponding
vertices.

However, if those counter terms do not own the expected structures or
the above expected relations do not hold,
we can necessarily conclude that the theory is not consistent with
the requirement of renormalizability, i.e. the theory is
non-renormalizable.

Before starting to extract those counter terms of vertices,
we write down (Here we use the Feynman and 't Hooft gauge
and work in the dimension regularization and ${\bar{MS}}$
renormalization scheme) the wavefunction renormalization
of $A^0$, $A^1$ and $A_5$.
\bea
Z_{A^0}&=& 1 + (N_{VB} \times \frac{ 10}{3}-{N_{S}\over3})\,\, C_{div},\\
Z_{A^1}&=& 1 +\frac{19}{3} \,\, C_{div}  ,\\
Z_{A_5}&=& 1 + 4\,\, C_{div}.
\eea
where $C_{div} = g^2 \kappa \Delta_{\epsilon} {C_2(G)}$.
The $N_{VB}$ is to
count the number of adjoint representation of
vector bosons and their ghosts, $N_{S}$ is to count
the number of adjoint representation of the scalar, and
in our case $N_{VB}=2$, $N_{S}=1$.
$C_2(G)$ is the Casimir operator of the
adjoint representation of gauge group G.
It is remarkable that the above result gives
$Z_{A^1}=Z_{A^0}$.

Now we start to construct the relevant counter
terms up to one-loop level through the corresponding five processes,
$A^0 \rightarrow A^0 A^0$, $A^0\rightarrow A^1 A^1$,
$A^0 A^0\rightarrow A^0 A^0$, $A^0 A^0\rightarrow A^1 A^1$,
and $A^1 A^1\rightarrow A^1 A^1$, respectively. The relevant
topologies of Feynman diagrams are given in Fig. 1. and
Fig. 2, respectively.

The counter terms of the relevant trilinear couplings are given below:
\bea
\delta Z_{000}&=&(N_{VB} \times {4 \over 3}-{N_{S}\over 3})\,\, C_{div} \,\, V_3\\
\delta Z_{011}&=&2 \times {4 \over 3}\,\, C_{div}\,\, V_3
+ {9 \over 2}\,\, C_{div} f^{abc} (R_2-1) \left [g^{\mu\nu} p^{\rho}-g^{\mu \rho} p^{\nu} \right]
\eea
$p$ is the incoming momentum of $A^0_{\mu}$. Then the renormalization
constant of the trilinear coupling of zero modes can be given as
\bea
Z_{000}=1+(N_{VB} \times {4 \over 3}-{N_{S}\over 3}) C_{div},
\eea

The counter terms of the relevant quartic couplings are given below:
\bea
\delta Z_{0000}& =&-( N_{VB} \times {2\over 3} +{N_S \over 3})\,\, C_{div}\,\, V_4,\\
\delta Z_{0011} &=&-{4\over 3}\,\,C_{div}\,\,V_4 +(R_2-1) T_{01},\\
\delta Z_{1111} &=&-{4\over 3}\,\,C_{div}\,\,V_4 + R_1^2 S_{11} + (R_2-1) T_{11} + (R_2^2-1) U_{11}.
\eea
where $T_{01}$, $S_{11}$, $T_{11}$ and $U_{11}$ are given as
\bea
T_{01}&=&{\kappa g^2 \Delta_{\epsilon}\over 4} \left \{ g^{\mu \nu} g^{\rho \sigma}\left[- {5\over 2} \Sigma^{ab}_{cd} C_2(G) - 5 S^{ab}_{cd}\right] \right .\nnb\\
&&\left .+g^{\mu \rho} g^{\sigma \nu} \left [ 4 f^{abe} f^{cde}  + 2 f^{eag} f^{gch} f^{hdi} f^{ibe}\right ] \right .\nnb\\
&&\left .+g^{\mu \sigma} g^{\rho \nu} \left [ -4 f^{abe} f^{cde} + 2 f^{eag} f^{gbh} f^{hci} f^{ide}\right ] \right \},\\
S_{11}&=&\kappa g^2 \Delta_{\epsilon} \left \{g^{\mu \nu} g^{\rho \sigma} \left[{1 \over 2}\Sigma^{ab}_{cd} C_2(G)+S^{ab}_{cd}\right ]
+g^{\mu \rho} g^{\nu \sigma} \left[{1 \over 2}\Sigma^{ac}_{bd} C_2(G)+S^{ac}_{bd}\right]\right .\nnb\\
&&\left .+g^{\mu \sigma} g^{\nu \rho} \left[{1 \over 2}\Sigma^{ad}_{bc} C_2(G)+S^{ad}_{bc}\right] \right \},\\
T_{11}&=&-{\kappa g^2 \Delta_{\epsilon}\over 4} \left \{g^{\mu \nu} g^{\rho \sigma} \left[23 \Sigma^{ab}_{cd} C_2(G)+30S^{ab}_{cd}\right]
+g^{\mu \rho} g^{\nu \sigma} \left[23 \Sigma^{ac}_{bd} C_2(G)+30 S^{ac}_{bd}\right]\right .\nnb\\
&&\left .+g^{\mu \sigma} g^{\nu \rho} \left[23 \Sigma^{ad}_{bc} C_2(G)+30 S^{ad}_{bc}\right] \right \},\\
U_{11}&=&\kappa g^2 \left \{g^{\mu \nu} g^{\rho \sigma} \left[{7\over 2} \Sigma^{ab}_{cd} C_2(G)+3 S^{ab}_{cd}\right]
+g^{\mu \rho} g^{\nu \sigma} \left[{7 \over 2} \Sigma^{ac}_{bd} C_2(G)+3 S^{ac}_{bd}\right]\right .\nnb\\
&&\left.+g^{\mu \sigma} g^{\nu \rho} \left[{7 \over 2} \Sigma^{ad}_{bc} C_2(G)+3 S^{ad}_{bc}\right] \right \},
\eea
where $S^{ab}_{cd}=f^{eaf} f^{fcg} f^{gbh} f^{hde}+ f^{eaf} f^{fdg} f^{gbh} f^{hce}$.
$S_{11}$ is the contribution of scalar $A_5^1$ in two-point one loop,
$T_{11}$ is from the three-point one-loop with one  $A^1A^1A^1A^1$ vertex,
and $U_{11}$ is from the diagrams with two $A^1A^1A^1A^1$ vertices.
And the convention of indices are given as
$A^{ia}_{\mu}\rightarrow A^{jb}_{\nu} A^{jc}_{\rho}$ and
$A^{ia}_{\mu} A^{ib}_{\nu}\rightarrow A^{jc}_{\rho} A^{jd}_{\sigma}$
Substituting $R_i$ into the sum of $R_1^2 S_{11}+(R_2-1) T_{11}+(R_2^2-1)U_{11}$
we get
\bea
\kappa g^2 \Delta_{\epsilon} \left \{g^{\mu \nu} g^{\rho \sigma} \left[{13\over 8} \Sigma^{ab}_{cd} C_2(G)+ {1\over 4} S^{ab}_{cd}\right]
+g^{\mu \rho} g^{\nu \sigma} \left[{13\over 8} \Sigma^{ac}_{bd} C_2(G)+ {1\over 4} S^{ac}_{bd}\right]\right .\nnb\\
\left.+g^{\mu \sigma} g^{\nu \rho} \left[{13\over 8} \Sigma^{ad}_{bc} C_2(G) + {1\over 4}S^{ad}_{bc}\right] \right \}.
\eea
So neither $\delta Z_{1111}$, nor $\delta Z_{0011}$,
nor $\delta Z_{011}$ have the expected structure.

The quartic coupling of the zero modes can be formulated as
\bea
Z_{0000}=1-(N_{VB} \times {2 \over 3} +{N_S \over 3}) \,\, C_{div},
\eea

The renormalizability of the zero modes part can be easily checked, since
the relation $Z_{000}^2=Z_{A^0} Z_{0000}$ indeed hold.
The non-renormalizability of the KK excitations is obvious from
the results given above. The difference of $Z_{000}$ and
$Z_{011}$ can be explained by two facts:
the first one is that there is no interaction
term of the form $\partial^{\mu} A_5 A^1_{\mu} A_5$, since
this term is forbidden by the requirement of the conservation
of the fifth momentum and is eliminated
in the procedure of integrating out the
fifth space. There is indeed one diagram in which $A_5$
contributes superficially divergently, but it is finite.
So the scalar contributes to the
$A^0 \rightarrow A^1 A^1$ convergently.
The second one is related
with the normalization factor of the
quartic interaction $A^1 A^1 A^1 A^1$, which provides
the terms related with the normalization factors $R_i$.
The differences between $\delta Z_{0000}$ and $\delta Z_{0011(1111)}$,
can also be explained by these two facts.

So, we see here that more than one counter terms are necessarily needed
in order to eliminate all ultraviolet divergences for the processes we
consider. In other words, the tree level
relations among couplings given by simply truncating
the infinite KK tower are not consistent with the requirement of
a renormalizable theory.
And it is in this sense that the simply truncated theory is
non-renormalizable.
As explained above, in the non-Abelian SU(N) gauge theory case,
it is the $R_i$ and the forbidden trilinear coupling
$\partial^{\mu} A_5 A^1_{\mu} A_5$ that conspire to make the
truncated theory non-renormalizable.
Therefore, in order to eliminate all divergences in the theory,
the more generic effective Lagrangian with one KK excitation
which respects the 4D Lorentz spacetime symmetry, the 4D zero
mode gauge symmetry
and the fifth momentum conservation law
should have the following form
\bea
L&=&-{1 \over 2} Tr[F_{\mu\nu} F^{\mu\nu}]
-{1 \over 2} Tr[{\bar F_{\mu\nu}} {\bar F^{\mu\nu}}]
- M_C^2 Tr[{\bar A_{\mu}} {\bar A^{\mu}}]
-\lambda_{21} Tr[{\bar A_{\mu}} D^{\mu} D^{\nu} {\bar A_{\nu}}]\nnb\\
&&-\lambda_{31} Tr[F_{\mu\nu} {\bar A^{\mu}} {\bar A^{\mu}}]\nnb\\
&&-\lambda_{41} Tr[{\bar A^{\mu}} {\bar A^{\nu}} ]
Tr[{\bar A_{\mu}} {\bar A_{\nu}}]
- \lambda_{42} Tr[{\bar A^{\mu}} {\bar A_{\mu}}]
Tr[{\bar A^{\nu}} {\bar A_{\nu}}]\nnb\\
&&-Tr[D^{\mu} A^1_5 D_{\mu} A^1_5]-M_C^2 Tr[A^1_5 A^1_5]
-\lambda_{33} Tr[{\bar A^{\mu} } {\bar A_{\mu}} A^1_5]\nnb\\
&&-\lambda_{43} Tr[{\bar A^{\mu} } {\bar A_{\mu}}] Tr[A^1_5 A^1_5]
-\lambda_{44} Tr[{\bar A^{\mu} } A^1_5] Tr[{\bar A_{\mu}} A^1_5]
-\lambda_{45} Tr[A^1_5 A^1_5 A^1_5 A^1_5]\nnb\\
&&+\cdots\,\,,
\label{efl}
\eea
where ${\bar F^{\mu\nu}}=D^{\mu} {\bar A^{\nu}} - D^{\nu} {\bar A^{\mu}}$,
$D^{\mu}=\partial^{\mu}-i g [A^{\mu}, . ]$,
${\bar A} = \sum_{a} {\bar A^a} T^a$, $T^a$ are the generators
of the gauge group, the Tr means to sum over the generators of the
gauge group, and the omitted terms are related with gauge fixing and
ghost terms. The effective Lagrangian is invariant under the
following transformation
\bea
A &\rightarrow& A'=U A U^{-1} - {i \over g} (\partial U) U^{-1}\nnb\\
{\bar A} &\rightarrow& {\bar A'} = U {\bar A} U^{-1}\nnb\\
A_5^1 &\rightarrow& A^{1\prime}_5 = U \bar A^1_5 U^{-1}\,\,.
\eea
After matching this generic effective Lagrangian with the truncated
RE4DT at the matching scale $\Lambda$, the ultraviolet boundary condition
of couplings $\lambda_{i}$ in Eq. (\ref{efl}) is fixed. Below
the matching scale $\Lambda$, these couplings will develop
in terms of their RGEs, respectively.

Compared the extra dimension model with the
renormalizable SU(5) unification model in 4D,
there is a similarity between these two theories:
the breaking of the tree level relations.
In the SU(5) unification model, the SM
is the effective theory of SU(5) GUT theory for energy scale below
the GUT scale $\Lambda_{GUT}$. At the $\Lambda_{GUT}$, there are
tree-level relations among the couplings of gauge
groups $SU(3)\times SU(2) \times U(1)$. Below the $\Lambda_{GUT}$,
due to the decoupling of Higgs multiplets and the SU(5) gauge
symmetry breaking, the gauge couplings develop respectively and
the tree level relations of them are broken by the quantum corrections.

There is a difference between these two theories: there are extra
operators in the extra dimension model generated by quantum corrections.
Compared with the renormalizable SU(5) where all
renormalizable terms of the subgroup
$SU(3)\times SU(2) \times U(1)$ have been contained in the Lagrangian
of SU(5) theory, the extra dimension SU(N) theory is unlucky in this respect.
Since the extra interaction terms, like $Tr[A^1_5 A^1_5 A^1_5 A^1_5]$,
although not permitted by the 5D Lorentz and 5D SU(N) gauge symmetry,
have to be introduced in order to remove divergences from
the theory.

In order to pinpoint the reasons for the breaking down of
tree level relations and the appearance of extra
operators, we consider the dimension reduction and rescaling procedure
of the renormalizable $\phi^4$ theory defined in 4D.
Assuming that the $z-$direction is compactified, by using the
dimension reduction and matching procedure, we will get its RE3DT
defined at a scale $\Lambda_{UV}^{3D}$. Since the RE3DT is a
super-renormalizable theory, vertex corrections are finite and
there is no need to introduce any a counterterm for the couplings
of the KK modes. However, after considering the quantum corrections,
the finite loop contributions still break the tree-level relations among
couplings of KK modes, the direct reason is still the different
normalization factor between zero mode and KK excitations.

Since in the simply truncated effective $\phi^4$ and SU(N) effective
theories, either in 4D or in 3D, the tree level relations among couplings
can not hold in the quantum corrections, although they are supposed to hold
in their underlying theories. The fundamental reason for the breaking down of
tree level relations seems to be related with the higher dimension
Lorentz symmetry and higher dimension gauge symmetry breaking, and
the dimension reduction and rescaling procedure itself.

We examined the truncated QED theory, where only the vector
boson is assumed to propagate in the bulk. The Lagrangian of the
theory in 5D has the form 
\bea
L&=&-{1\over 4} F^{MN} F_{MN} + {\bar \psi}(i \gamma^{\mu} D_{\mu}-m) \psi \delta(x_5)
\eea
where $F_{MN}=\partial_{M} A_N - \partial_{N} A_M, \,\, M=0,1,2,3,5$,
$\mu=0,1,2,3$, $\psi$ is defined in the 3-brane and $D_{\mu}\psi =\partial_{\mu} \psi- i g A_{\mu} \psi$. This theory is non-renormalizable in 5D
due to the fact that the gauge coupling constant $g$ has a negative
mass dimension. We find that up to one-loop
level, the tree level coupling structure is unchanged by the
quantum corrections.
The reason seems to be simple: the bilinear interaction vertices and 
normalization factors of the theories
do not undermine the tree level relations among couplings in
these two cases, not as in the 
$\phi^4$ and non-Abelian SU(N) gauge theories where the normalization
factors of quartic couplings or forbidden terms break down
the tree level relations.
We also examined the real scalar $\phi^3$ theory in 5D, and this theory is
super-renormalizable according to the power law. The Lagrangian is given
by
\bea
L={1\over2} (\partial_{M} \phi_{5D}) (\partial^{M} \phi_{5D})
-{1\over 2}m^2 (\phi_{5D})^2 - {\lambda_{5D} \over 3!} (\phi_{5D})^3.
\eea
And again we find that up to one-loop level, the coupling structure
of its truncated theory is unchanged by the quantum corrections.

In summary, up to one loop level, by truncating KK excitations to
only one, we examined the renormalization of the
truncated KK theories of $\phi^4$ theory, the non-Abelian gauge
SU(N) theory, QED theory, and $\phi^3$ theory defined in 5D,
and found that the normalization factors
of four KK excitations, or the forbidden missing terms, or both
undermine the tree-level structure of the
simply truncated theories in quantum corrections.
We conclude that the breaking of the higher dimension Lorentz symmetry and
higher dimension gauge symmetry, interactions assumed in the underlying
Lagrangians, and the dimension reduction and rescaling procedure
play their roles in breaking down of the tree level relations.


\acknowledgements
One of the authors, Q.S. Yan, would like to thank
Prof. Yu Ping Kuang, Dr. Tianjun Li, Prof. Zhong Qi Ma, and
Prof. Qing Wang for helpful discussions. To analyze divergences and
construct counter terms, we used the FeynArts and FeynCalc 4.1.0.3b
(http://www.feyncalc.org). This work is supported by National Natural
Science Foundation of China and State Commission of Science and
Technology of China.

\newpage
\begin{center}
{\large \bf Figure Captions}
\end{center}
\begin{figure}[h]
\begin{center}
\epsfysize=4truecm\epsfbox{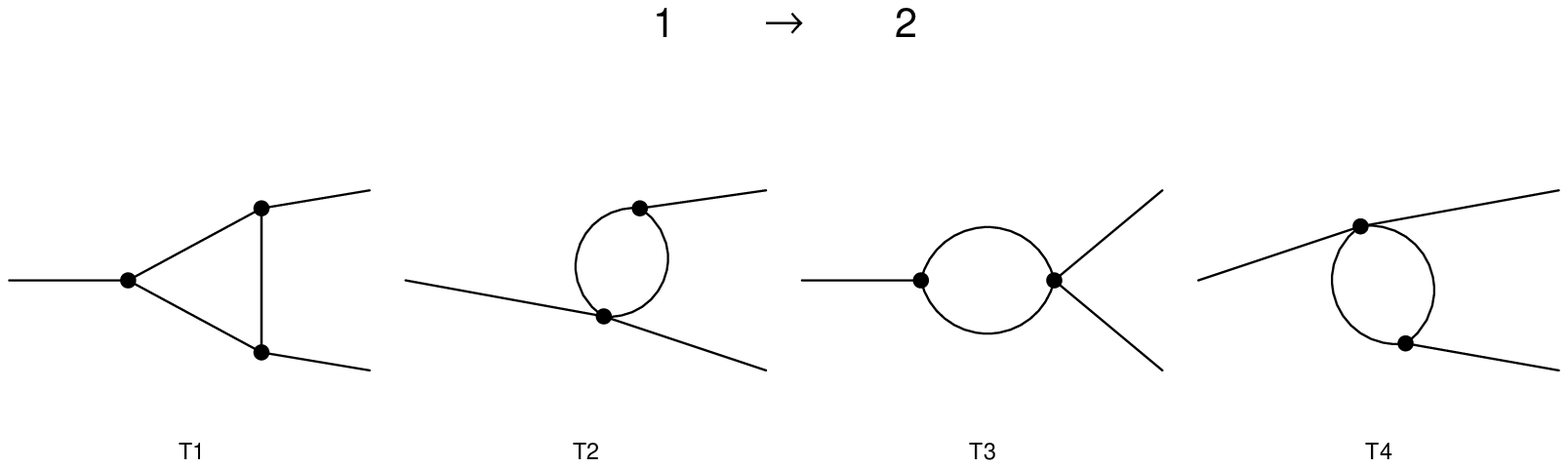}
\vspace{0truecm}
\end{center}
\caption{The topologies of $1\rightarrow 2$ processes}
\label{fig1}
\end{figure}

\begin{figure}[h]
\begin{center}
\epsfysize=11truecm\epsfbox{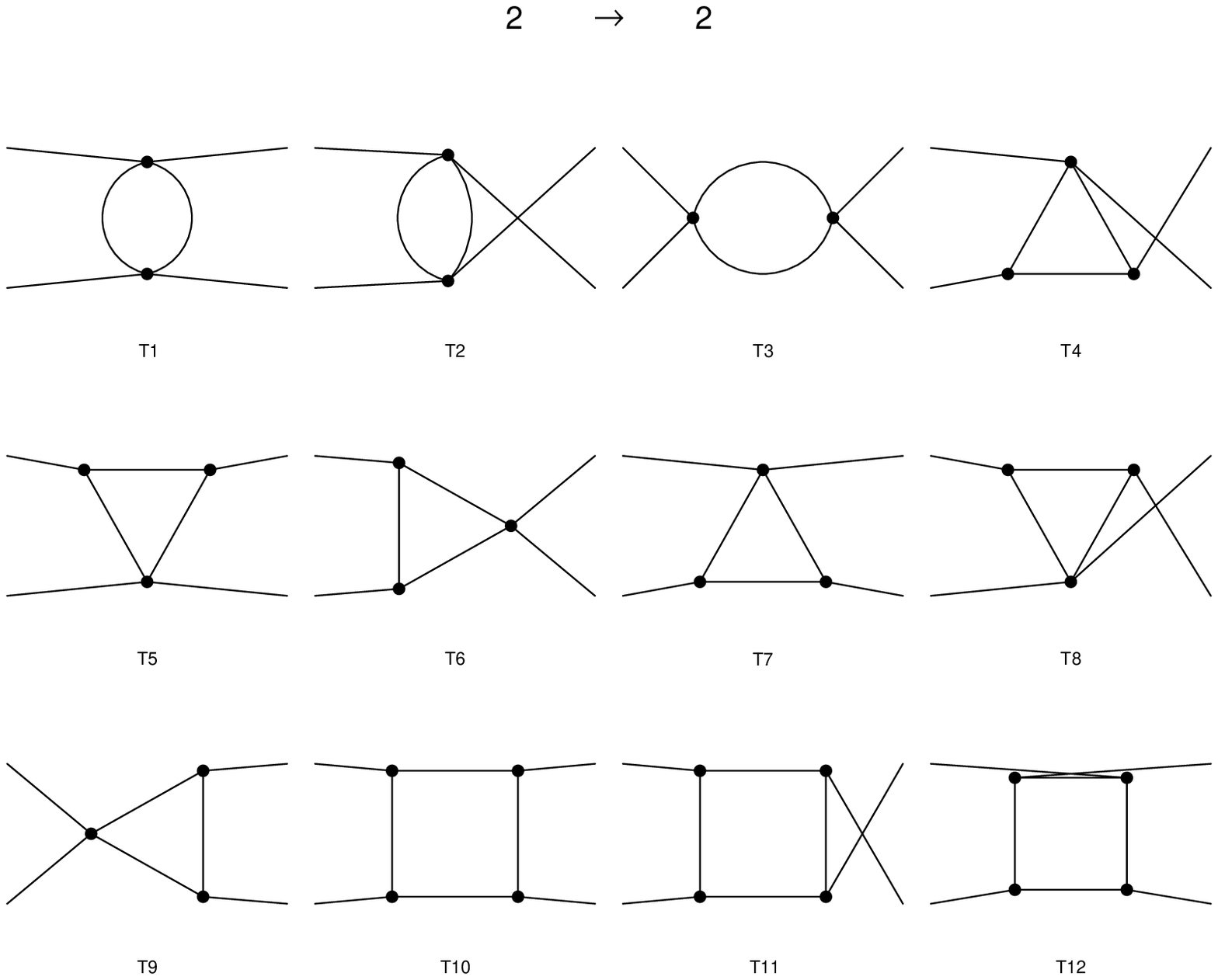}
\vspace{0truecm}
\end{center}
\caption{The topologies of $2\rightarrow 2$ processes}
\label{fig2}
\end{figure}

\end{document}